 \documentclass[aps,reprint,superscriptaddress]{revtex4-1}

\usepackage{graphicx}
\usepackage{amsmath}
\usepackage{amssymb}
\usepackage{bm}
\usepackage{cancel}
\usepackage[normalem]{ulem}
\usepackage{xfrac}
\usepackage[colorlinks=true,linkcolor=blue,citecolor=blue,urlcolor=blue]{hyperref}

\usepackage{textcomp}
\usepackage{color}
\usepackage{xcolor}

\usepackage{hyperref}
\usepackage{cleveref}

\usepackage{subfigure}
\usepackage{placeins}
\usepackage{siunitx}
\usepackage{blindtext}

\begin{document}

\title{Topologically controlled emergent dynamics in flow networks}

\author{Miguel Ruiz-Garc\'ia}

\affiliation{Department of Physics and Astronomy, University of Pennsylvania, Philadelphia, PA 19104, USA}

\author{Eleni Katifori}

\affiliation{Department of Physics and Astronomy, University of Pennsylvania, Philadelphia, PA 19104, USA}

\date{\today}

\begin{abstract}
Flow networks are essential for both living organisms and enginneered systems. These networks often present complex dynamics controlled, at least in part, by their topology. Previous works have shown that topologically complex networks interconnecting explicitly oscillatory or excitable elements can display rich emerging dynamics.  Here we present a model for complex flow networks with non-linear conductance that allows for internal accumulation/depletion of volume, without any inherent oscillatory or excitable behavior at the nodes. In the absence of any time dependence in the pressure input and output we observe emerging dynamics in the form of self-sustained waves, which travel through the system. 
The frequency of these waves depends strongly on the network architecture and it can be explained with a topological metric.
\end{abstract}
\maketitle

Flow networks are necessary when nature or humans need to distribute resources, such as water, nutrients or power, to distances larger than what diffusion alone would allow. The human circulatory system, the interconnected networks of fungal hyphae, or the power distribution networks that provide electricity to households are perhaps some of the most well known examples of webs that efficiently transport resources. The network topology determines properties such as the robustness, efficiency, and cost of the distribution system. As a result, the interplay between flow network topology and function has been explored in many different contexts~\cite{LaBarbera1990,Bebber2007,Katifori2010a,Chang2017,Gavrilchenko2018}. On the other hand, artificial microfluidic networks are rapidly evolving, incorporating more elaborated topologies and non-linear elements that will enable a complete tuned-in control of the internal flows~\cite{stone2009tuned-in,duncan2013pneumatic,case2019braess}, eliminating or reducing to the minimum the need for external hardware.

Systems modeled by flow networks are seldom static --their function is typically dominated by fluctuations that lead to rich dynamics. These fluctuations can be extrinsically driven by time-varying environmental stimuli --like blood pulses produced by the heart. But they can also be spontaneous, i.e. manifest even in the absence of a clear fluctuating external stimulus. This type of spontaneous oscillations has been observed in physarum where it has been explained as a manifestation of peristalsis in a random network~\cite{alim2013random}, and putatively in the mammalian cortex, in the form of unexplained hemodynamic oscillations that do not directly correlate with neuronal function~\cite{winder2017weak}.

Inspired by such systems, in this work we explore to what extent oscillations or fluctuations in flow networks of an arbitrary topology can emerge spontaneously, in the absence of a time-modulating external stimulus, as a product of internal non-linearities. Models that can produce complex spontaneous dynamics already exist in the network context, for example the broad class of Kuramoto models of coupled oscillators~\cite{Niebur1991,Laing2016,Wetzel2017}. However, in these models, oscillatory dynamics is already explicitly assumed, and the global dynamics results from synchronization between the nodes. In this work we show that a minimal, but still physical model of non-linear conductance can, for a broad range of parameters, produce spontaneous self-sustained oscillations. This excitation phenomenon has a global nature, genuinely different from other models of excitable networks that use interconnected elements of an explicitly excitable nature (e.g. \cite{kinouchi2006optimal,roxyn2004selfsustained}).

Additionally, we show how the frequency and amplitude of the oscillations depends on the network structure, and that this dependence can be explained using a simple network metric. Network topology has previously been connected to dynamics in other models of complex networks, e.g. see~\cite{larremore2011predicting,castellano2017relating,miguel2018effects}, but to our knowledge, not in the context of systems of non-linear resistors and their oscillations. In what follows we adopt the language and terminology of fluid flow networks, although the results can be also expressed in the electrical network context. In particular, a variation of the $1$D limit of this model has been previously used to study the dynamics of semiconductor superlattices \cite{bonilla2005non,bonilla2010nonlinear}, which also present self-sustained oscillations.

\begin{figure}
\includegraphics[width=0.95\linewidth]{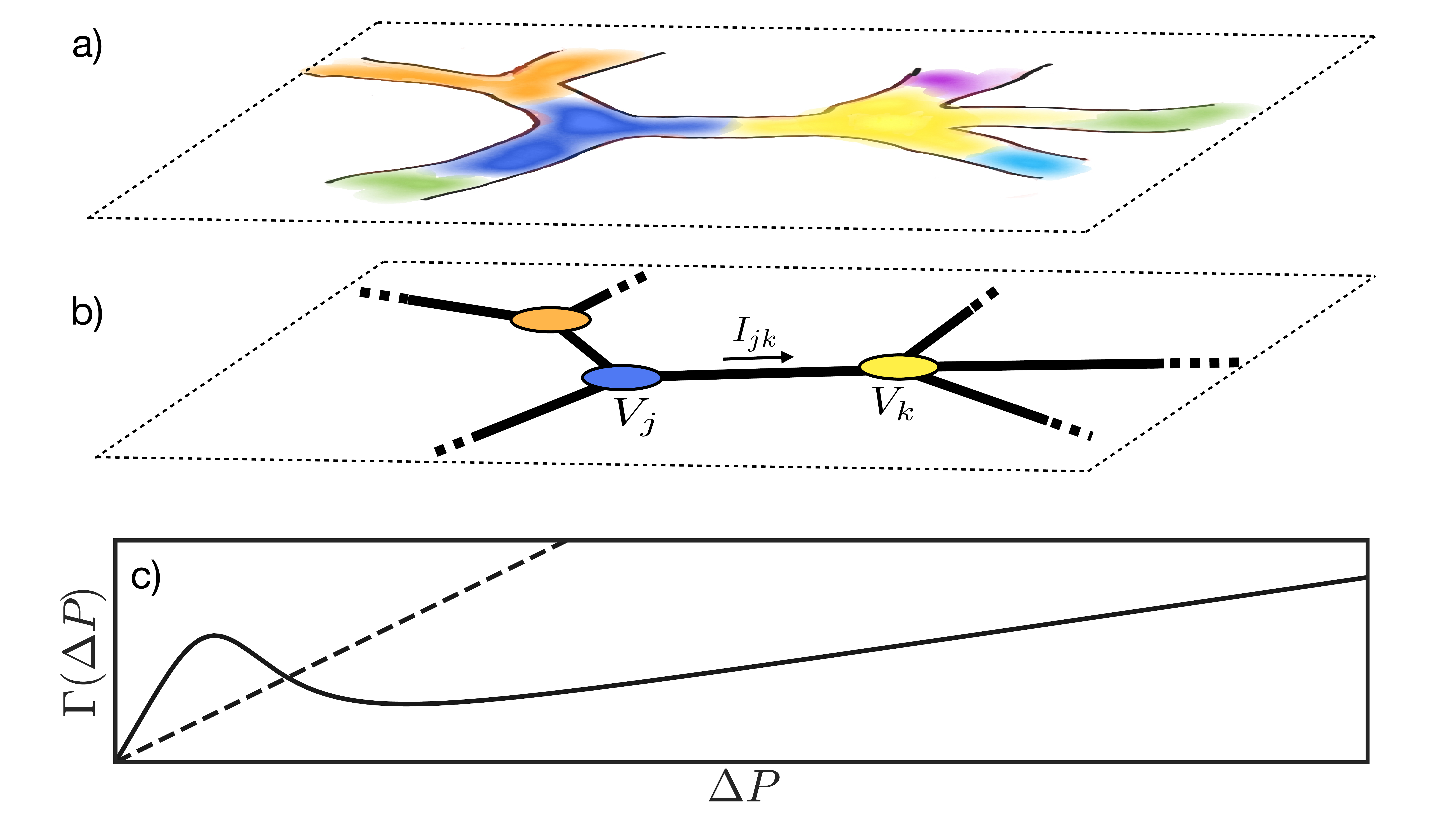}
\caption{ Modeling non-linear flow networks. (a) Illustration of a small portion of a physical flow network. Colors separate the volume that surrounds each node. (b) The idealized version of (a) that we use in the model. The volume contained within the surroundings of every node $k$ in (a) it is now accounted at the node as $V_k$. Currents $I_{jk}$ flow between the nodes. (c) Examples of an Ohmic or linear (dashed) and a non-linear (continuous line) function $\Gamma(\Delta P)$ that control the current between nodes.}
\label{fig1}
\end{figure}

We define a network as a set of $N$ nodes and connections between them (edges). Pressures ($P_i$) and accumulated volumes ($V_i$) are defined at each node $i$ and can be time dependent. The volumetric current $I_{ij}$ is defined as the current from node $i$ to node $j$ on edge $ij$, as in figure \ref{fig1}. Unless otherwise stated, we randomly choose $n$ nodes from the network and externally control their pressure, setting the inputs and outputs for the system (analogous to connecting a battery to a resistor network). Without loss of generality, in what follows we will use non-dimensional fields for $I$, $P$, $V$, and the time $t$. 

The current $I_{ij}$ depends on the pressure difference $\Delta P_{ij} = P_j-P_i$. In a simple Ohmic network this relationship would be linear, but in this work we consider the following general pressure flow relation:
\begin{equation}
I_{ij} = 
    \begin{cases}  \frac{1}{2}
    V_i^2 \Gamma(\Delta P_{ij}) , \quad \text{if } P_i>P_j\\
     \frac{1}{2} V_j^2 \Gamma(\Delta P_{ij}) , \quad \text{if } P_j>P_i
    \end{cases}
    \label{eq_I}
\end{equation}
where the current also depends on the accumulated volume at the node from where it flows, recovering the right scaling for Poiseuille flow for regions of $\Delta P_{ij}$ where $\Gamma(\Delta P_{ij})$ behaves linearly, see also \cite{PRE}. In this work, $\Gamma$ can display two different behaviors, either linear:
    \begin{equation}
    \Gamma_{L}(\Delta P) = h \Delta P,  
    \label{eq_gammas_l}
    \end{equation}
or nonlinear:
    \begin{equation}
    \Gamma_{NL}(\Delta P) = 
      \frac{1 + \epsilon  \Delta P^4}{1+\Delta P^4}\Delta P, 
    \label{eq_gammas_nl}
    \end{equation}
where $\epsilon$ and $h$ are dimensionless parameters that determine the shape of the non-linearity in $\Gamma_{NL}$ and the slope of $\Gamma_L$. These two different kinds of edges are compared in Fig.~\ref{fig1}(c). We choose a $\Gamma_{NL}$ such that it has a negative slope region. This makes stationary solutions with a homogeneous pressure drop in that range to be unstable (see \cite{PRE} for more details). Conservation of mass imposes the temporal variation of the accumulated volume at every node,
\begin{equation}
    \dot{V_i} = \sum_{j} I_{ij}.
    \label{eqCons}
\end{equation}
Finally, we include a constitutive relation that couples the excess volume from a dimensionless baseline (set to 1), to the curvature of the pressure:
\begin{equation}
    V_i -1 = \alpha \sum_j L_{ij} P_j.
    \label{eq_DV}
\end{equation}
$L_{ij}$ is the $ij$ element of the graph Laplacian $L=D-A$, where $D$ is the degree matrix defined as $D_{ij}=d_i \delta_{ij}$ with $d_i$ the degree of node $i$, and $A$ is the adjacency matrix. Note that if $V_i=1$ for every node except for node $k$ where $V_k>1$, relation \eqref{eq_DV} will create a pressure field that induces currents that disperse the accumulated volume at node $k$. If $V_k<1$ then the pressure field will create currents that restore the volume at node $k$. The exact constitutive relation \eqref{eq_DV} will generally depend on the specific properties of the system being modeled. 

From equations \eqref{eqCons} and \eqref{eq_DV}  we get:
\begin{equation}
     \alpha \sum_j L_{ij} \dot{P}_j =  \sum_j {I}_{ij},
    \label{eqdotP}
\end{equation}
which is the system of ordinary nonlinear differential equations that, together with the boundary conditions (externally set pressures) is propagated in time to produce the results presented in this work. See \cite{PRE} for more details.

\begin{figure}
\includegraphics[width=0.95\linewidth]{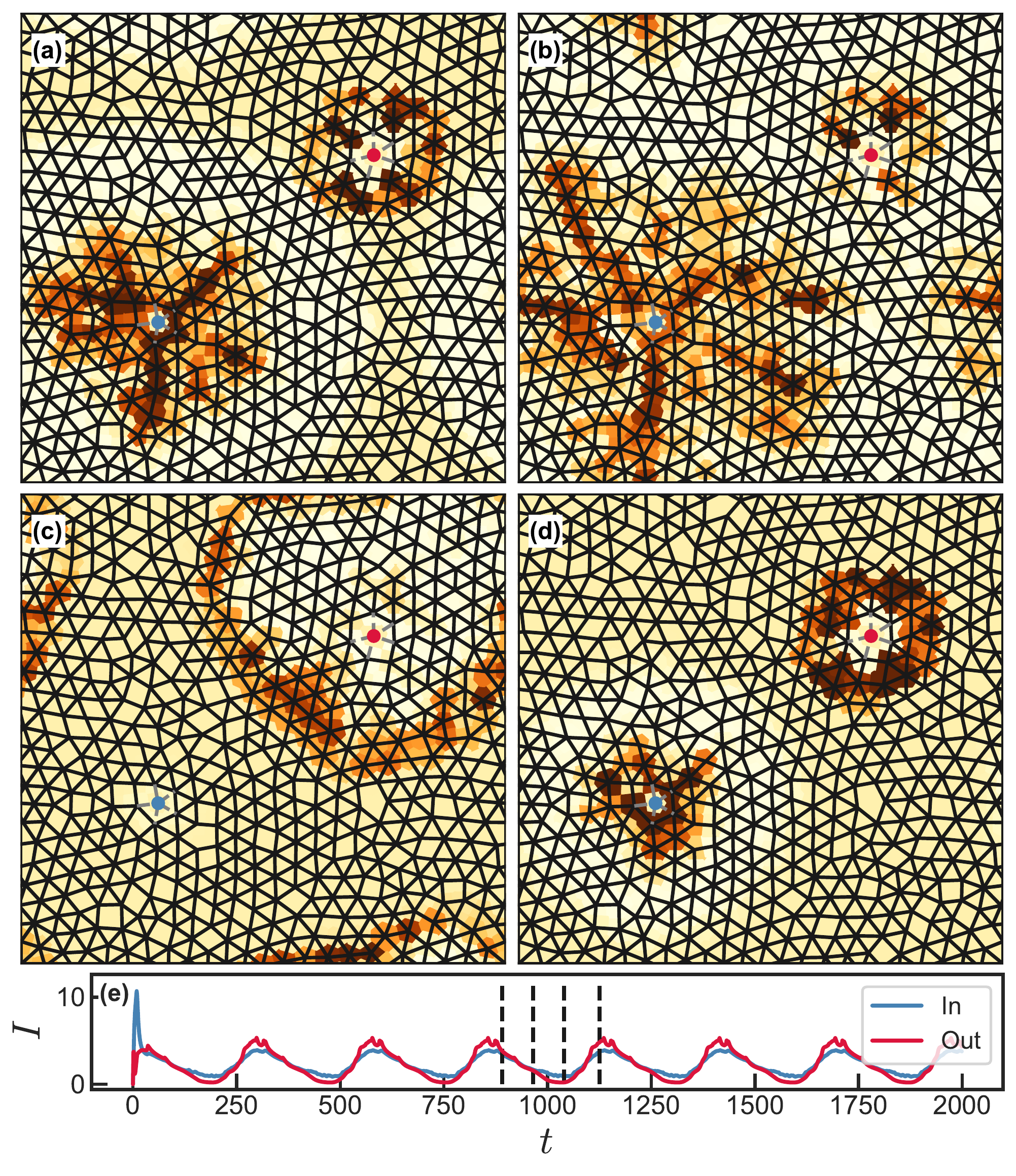}
\includegraphics[width=0.95\linewidth]{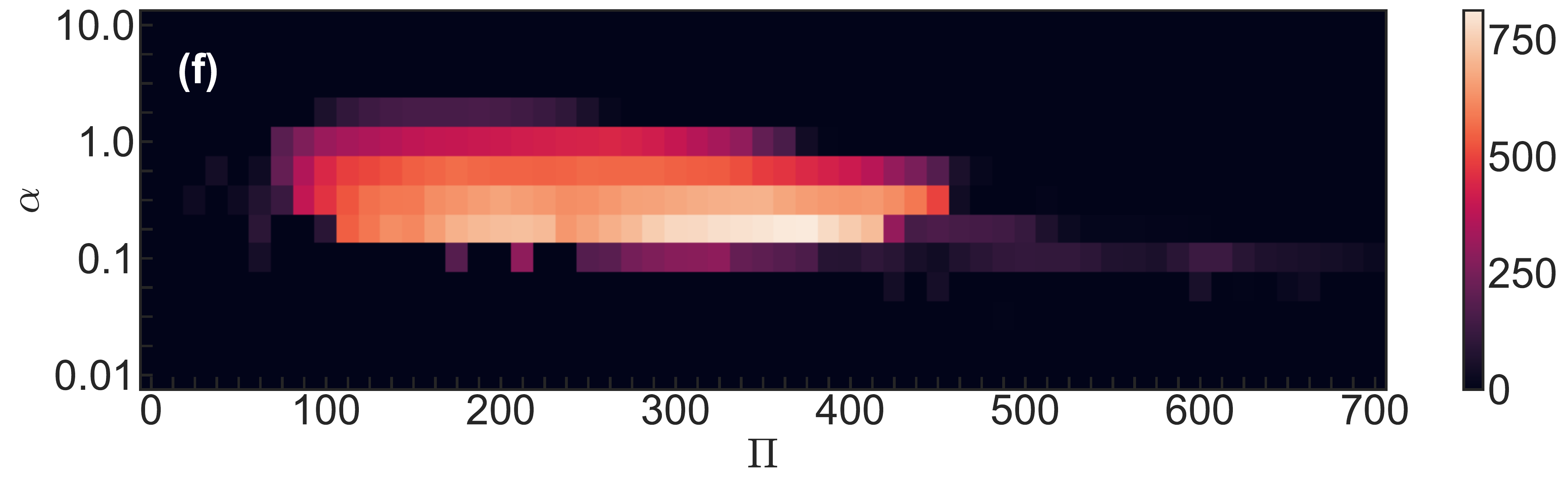}
\caption{Self-sustained waves in a disordered planar network with two contact points. All edges follow $\Gamma_{NL}$ (continuous lines) except for the ones connected to each of the contacts that follow $\Gamma_{L}$ (gray dashed lines). The network has $512$ nodes. (a)-(d) Snapshots for different simulation times. To aid visualization we color the area around each node according to the volume at the node ($V_i$), color scheme goes from $V=0$ (light yellow) to $V=6$ (dark brown). Pressure is imposed at the contact points (blue and red dots). (e) Total current going in and out of the system versus time, vertical dashed lines mark the time at which each snapshot (a)-(d) was taken. Initial conditions were $P_i=0$ and $V_i=1$ for every node. (f) Phase diagram presenting the magnitude of the oscillations $\sum_i \langle (P_i(t) - \langle P_i(t) \rangle )^2 \rangle $ for different values of $\alpha$ and $\Pi$.}
\label{fig2}
\end{figure}

Although the values for $\epsilon$ and $h$ can quantitatively affect the behavior of the system, spontaneous dynamics appears for a wide range of these parameters. 
In this work we focus on the dependence of the dynamics on network topology, so we fix $\epsilon$ and $h$ to $\epsilon=0.001$ and $h=1/5$.  For all simulations the initial conditions are $V_i=1$ and $P_i=0$ for all $i$. The pressure is externally imposed throughout the simulation at a set of points that we term contact points. The red contact point in figure \ref{fig2} has fixed pressure zero, whereas the pressure at the blue contact point linearly increases from $0$ until it reaches $\Pi>0$ at $t=t_0$ and is kept constant afterwards.

Fig. \ref{fig2} shows a simulation for a disordered planar network with periodic spatial boundary conditions and an average coordination degree of $\sim 5.5$. Panels (a)-(d) show the spatial distribution of accumulated volume for different snapshots of the simulation. For these panels we use $\alpha=0.32$, $\Pi=150$ and $t_0=30$. The accumulated volume displays self-sustained oscillations in the form of pulses that travel from the high to the low pressure contact point. Fig.~\ref{fig2} (e) shows the net current that is going in and out of the system at the contact points as a function of time. In particular both currents are in phase, showing that when one pulse is leaving the system another one is entering. Note that both currents are not identical, therefore the total volume of the system is not conserved. The dashed lines mark the times corresponding to panels (a)-(d). In Fig.~\ref{fig2} (f) we plot the sum of the variance of the pressure $P_i(t)$ at each node as a proxy for spontaneous dynamics. Black regions in panel (f) correspond to stationary behavior whereas the brighter colors represent larger oscillations. Oscillations appear when the pressure difference $\Pi$ between the contacts is large enough so that the negative slope region of $\Gamma_{NL}$ is probed (Fig. \ref{fig1} (c)). 
These oscillations persist when we modify the network properties or contact point number and location. In what follows we explore how changing the topology of the network, the number and distance between the contacts, and the distribution of linear and non-linear edges, affect the results.

\begin{figure}
\includegraphics[width=0.95\linewidth]{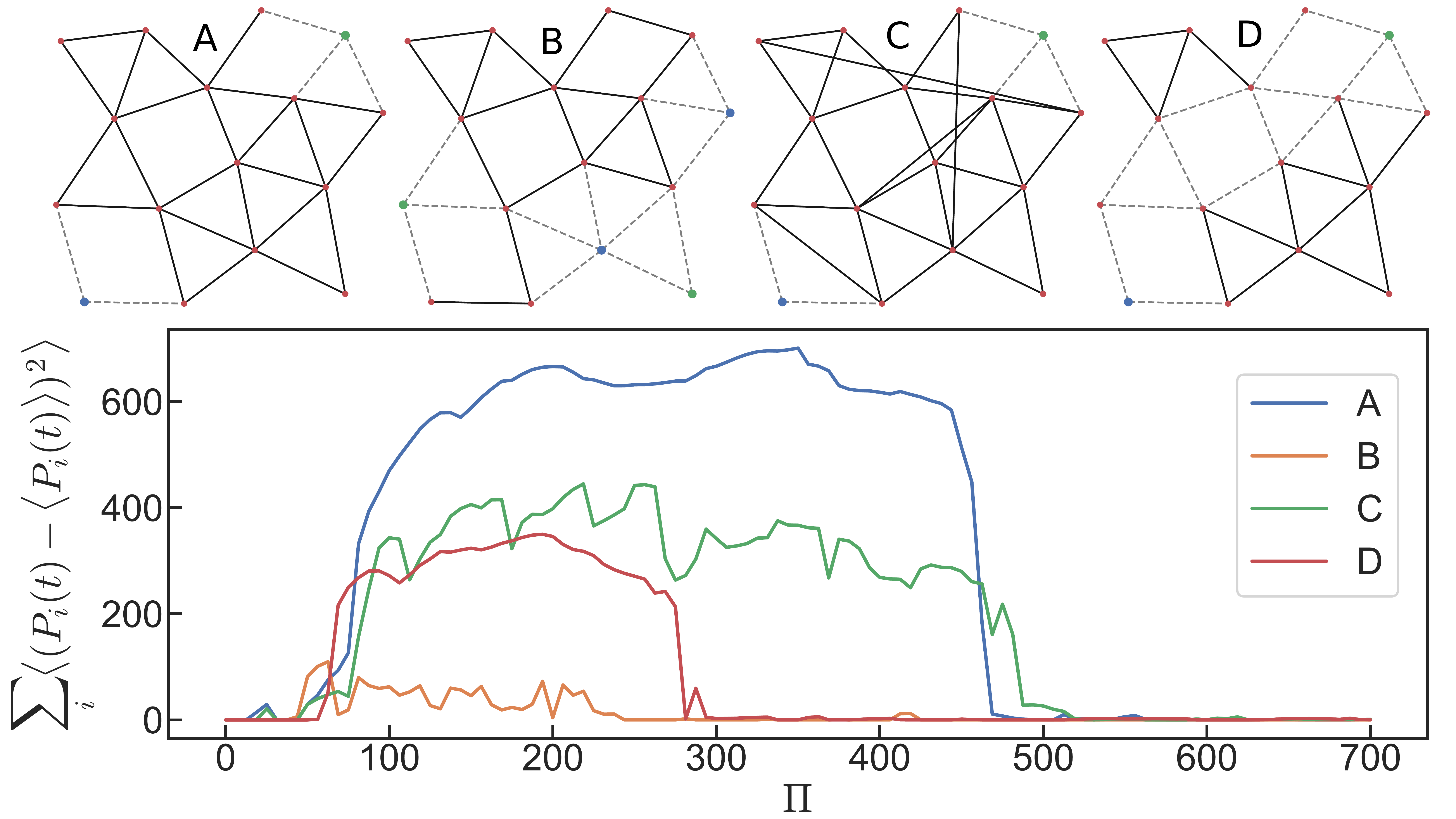}
\caption{Magnitude of the fluctuations versus external applied pressure for different network structural modifications of the planar network of Fig.~\ref{fig2} for $\alpha = 0.32$. For visualization in the top row we show  sketches of networks corresponding to each modification performed to the network showed in Fig. \ref{fig2}: (A) Two pressure contacts (same configuration as in Fig. \ref{fig2}), (B) $5$ high pressure and $5$ low pressure contacts at random positions, (C) addition of $26$ randomly distributed shortcuts, and (D) replacement of $360$ non-linear edges (following $\Gamma_{NL}$) chosen at random with linear edges (following $\Gamma_{L}$).
Bottom panel shows the magnitude of the fluctuations for each of the modifications.}
\label{fig3}
\end{figure}

In Fig. \ref{fig3} we demonstrate the existence of spontaneous dynamics after modifying the structure of the network. As a baseline, we use the same network as in Fig. \ref{fig2}, setting $\alpha = 0.32$, a value chosen as it corresponds to a dynamically rich region of the phase diagram in Fig. \ref{fig2} (e). We consider three different types of modifications of the system, represented pictorially by small networks sketches $B-D$. The original system is denoted with A in Fig. \ref{fig3} (blue line). In modification B, we increase the number of contact points to a total of $10$, randomly distributed (orange line). In modification C we include shortcuts between nodes chosen at random (green line), and finally, in modification D we replace a randomly selected fraction of non-linear edges with linear ones (red line). As Fig. \ref{fig3} indicates, in all cases, spontaneous generation of self-sustained oscillations persists for a broad region of $\Pi$, but the shape of the phase diagram is altered, at least for this value of $\alpha$. These particular cases illustrate the robustness of the oscillations to drastic changes to the system. We can observe that the amplitude of the oscillations is greatly reduced for case B, indicating that reducing the distance between the contacts has a strong effect on the dynamics.

\begin{figure}
\includegraphics[width=0.95 \linewidth]{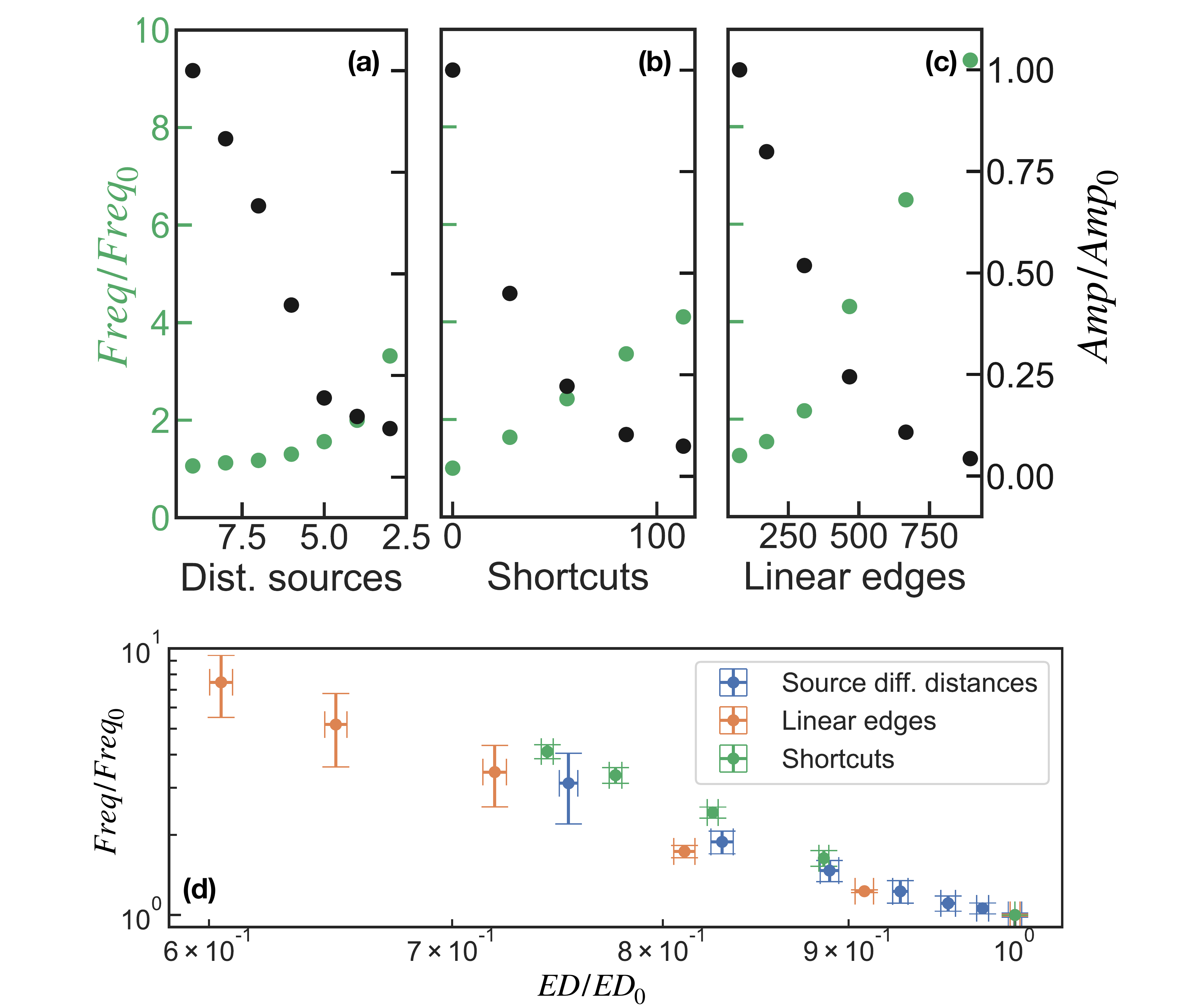}
\caption{Amplitude and frequency of the oscillations for different modifications to the network shown in Fig. \ref{fig2}. We use $\Pi=212.5$ and $\alpha = 0.32$.
(a) The two contacts are randomly placed on the planar network at a specified distance.
(b) Contacts are at a fixed distance whereas the topology of the network is changed adding shortcuts, in the form of new edges (following $\Gamma_{NL}$) between nodes chosen at random (excluding the two sources).
(c) Contacts are fixed whereas we modify the nature of the edges in the network, by replacing a set of edges chosen at random with linear edges. 
(d) All frequency points collapse on a single curve when plotted against the normalized effective distance (ED), inversely proportional to the escape probability of a random walk.
Error bars correspond to the standard error of the mean of the frequency distribution.}
\label{fig4}
\end{figure}

In Fig.~\ref{fig4} we study how the nature of the spontaneously emerging oscillations changes as the network becomes more interconnected, as the distance between the contact points decreases, or as we increase the fraction of linear edges. Again, we use the network described in Fig. \ref{fig2} (with two contacts) as the baseline. We fix the steady state pressure between the contacts to $\Pi=212.5$ and $\alpha=0.32$. The top panels of Fig. \ref{fig4} show how the amplitude and frequency of the oscillations change as we transform the baseline network: in panel (a) we show what is the effect of changing the distance between the two contacts, in panel (b) we keep the contacts in the same position but we add an increasing number of shortcuts, and finally, in panel (c) we randomly replace nonlinear edges with their linear counterpart. For every point in each of the three panels we carried out $100$ simulations. To characterize the dynamics of each simulation we compute the total accumulated volume in the system as a function of time. The total accumulated volume is calculated by the time integral of the difference of the total incoming and outgoing currents at the contact points, analogous to Fig. \ref{fig2} (e), and it is an oscillatory function. In Fig. \ref{fig2} (a)-(d), a pulse exits the system at the same time another one is created. The time that it takes for the pulse to traverse the system sets the frequency of the oscillations. After discarding the initial transitory period, we measure the amplitude of the oscillations of the total accumulated volume in the system. To measure their frequency, we compute the Fourier transform of the oscillations and record the value corresponding to the largest peak. In Fig.~\ref{fig4}(a), (b) and (c) we show the mean of the frequency and amplitude for all the simulations showing self-sustained oscillations. Note that under three completely different modifications of the network structure, the properties of the oscillations follow a similar general trend: an increasing frequency and decreasing amplitude. 

This trend can be characterized using a unified topological metric that explains the change in frequency as a response to the three structural changes in all three modifications. As the pulses travel from one contact to the other, the time that they take to travel through the system will decrease as the distance between the contacts is reduced, making the frequency go up. We therefore propose an \textit{effective network distance} that unifies all the structural modifications. This a purely topological metric that characterizes the structure of the network. To compute it we resort to the escape probability of a random walk (RW)~\cite{doyle2000random,redner2009fractal}. This is the probability that a RW starting from one contact arrives to the other contact without returning to the initial position. When the RW is on one node, we consider an equal probability to jump to any of the neighboring nodes. It is easy to understand that the escape probability increases as the two contacts are moved closer to each other. Likewise, adding more shortcuts increases the number of alternative paths between the two contacts, this way also increasing the probability that the RW will escape. Finally, as we observe numerically (see \cite{PRE}), linear resistors result in the pulse travelling much faster through them. To account for this in calculating the effective distance between the contacts, through the escape probability, linear edges are assigned a probability that the RW will traverse them that is much higher than the one assigned to non-linear edges. Therefore, when the RW lands on a node connected to one or more linear edges it will not traverse the non-linear ones in the next step. Instead, it will use, with equal probability, one of the linear edges (see Supplementary Materials for more details on how to compute the escape probability). In the bottom panel of Fig. \ref{fig4} we plot all the frequency data points presented in the top three panels, normalized by the frequency of the unaltered system (the simulation shown in Fig. \ref{fig2}). On the horizontal axis we plot the inverse of the escape probability for every configuration -- which we term effective distance --  averaged for each group of simulations. Once normalized by the effective distance of the unaltered system, we can plot all the points resulting from the three different modifications of the network as a function of the same topological metric. Finally, the fact that all the points on Fig. \ref{fig4} (d) lay on the same curve suggests that the effective distance is a meaningful measure to characterize an important part of the emerging dynamics in this system.

In summary, in this work we have shown that a network of nonlinear resistors can support emerging dynamics even in the absence of a time varying input. Depending on the value of the model parameters, the model displays stationary behavior or stable oscillations. The self-sustained oscillations are a robust phenomenon and persists for a broad range of graph topologies and choices for the contact points. Moreover, the dynamical behavior shown in this work does not depend on the specifics of the pressure current relation. For example, we also see self-sustained oscillations if we use $I_{ij} \propto V_i \Gamma(\Delta P_{ij})$ (note that $V_i$ is not squared here). Similarly, other functional forms for $\Gamma_{NL}$ also produce qualitatively analogous results, as long as there is a negative-slope region.

We have further shown how the structure of the network can affect the properties of emerging spontaneous fluctuations. In particular, the network topology and the amount of non-linear edges control the amplitude and frequency of the oscillations. Furthermore, the relation between the frequency of the oscillations and the structure of the network can be explained through a simple network metric.

This work was presented in the context of an incompressible fluid flowing in flexible tubes. However, our model and the results of this work are applicable to a broad class of physical networked systems where: (i) There is a conservation of some quantity (equation \eqref{eqCons} in our model), including possible accumulation/depletion of it throughout the system, a property that can also apply to compressible fluids in rigid pipes or electrons in semiconductors; (ii) there is a non-linear response of the flow to the change on the field that is driving the transport ($\Gamma_{NL}(\Delta P)$ in our case), and, (iii) there is a relation that couples the field that drives the flow to the amount of substance being transported (equation \eqref{eq_DV} in our model).

The system presented in this paper is a new model for excitable networks of arbitrary topology that to our knowledge has not been discussed before in the literature. Previous models for excitability in complex networks used expressions for already excitable elements that were connected within a complex network. In particular, these elements broadly belonged to two different classes: they were discrete variables that could be in a resting, excited or refractory state and that could excite their neighbors, e.g.~\cite{kinouchi2006optimal}; or they could be neuron-like continuous variables whose dynamics were coupled to their neighbors' dynamics, e.g.~\cite{roxyn2004selfsustained}. In our work the nodes are not intrinsically excitable, but simply store volume. Excitability emerges as a global effect that stems from the combination of: (i) the coupling between the stored volume and the pressure field and (ii) the nonlinear conductance of the edges. This combination gives rise to the complex dynamics shown, at least in part, in this work.

\begin{acknowledgements} 
This research was supported
by the National Science Foundation via Award No. DMR1506625 (M.R.G.), and the Simons Foundation via Award No. 454945 (M.R.G.). EK acknowledges support by NSF Award PHY-1554887, IOS-1856587, the University of Pennsylvania Materials Research Science and Engineering Center (MRSEC) through Award DMR-1720530, the University of Pennsylvania CEMB through Award CMMI-1548571, and the Simons Foundation through Award 568888. 
\end{acknowledgements}

 \bibliographystyle{apsrev4-1} 

\bibliography{bibliography.bib}

\end{document}